\documentstyle[preprint,aps]{revtex} % PREPRINT format
\begin{document}
\draft
%**end of header
\title{Density-Polarization Functional Theory of the response of
a periodic insulating solid to an electric field}
\author{X. Gonze$^{\star}$, Ph. Ghosez$^{\star}$, and R.W.
Godby$^{\dagger}$}
\address{$^{\star}$ Unit\'e PCPM,
Universit\'e Catholique de Louvain,
B-1348 Louvain-la-Neuve, Belgium}
\address{$^{\dagger}$ Cavendish Laboratory, University of Cambridge,
Madingley Rd.,
Cambridge CB3 0HE, United Kingdom}
\date{\today}
\maketitle
\begin{abstract}
The response of an infinite, periodic, insulating, solid to an
infinitesimally
small electric field is investigated in the framework of Density
Functional Theory. We find that the applied perturbing potential is not
a unique
functional of the periodic density change~: it depends also on
the change in the macroscopic {\em polarization}.
Moreover, the dependence of the exchange-correlation
energy on polarization induces an exchange-correlation
electric field. These effects are exhibited for a model
semiconductor. We also show that the scissor-operator technique
is an approximate way of bypassing this polarization dependence.
\end{abstract}
%
%\vskip 5mm
\pacs{PACS numbers: 71.10.+x; 77.20.+y; 78.20.Bh}
%\begin{multicols}{2}[]
\newpage

Density-Functional Theory (DFT)\cite{Hohenberg64,Kohn65}
has become the standard method
for first-principles calculations of the properties of solids.
Within the Local Density Approximation (LDA)\cite{Kohn65},
the accuracy obtained for
a wide set of properties is of the order of a few percent
\cite{DFTReviews}.
Notable exceptions to this gratifying picture are the cohesive energy
of solids, the energy gap of semiconductors, and their dielectric
susceptibility.
The first of these failures is attributed to the LDA
\cite{DFTReviews}, while the second comes from
a discontinuity in the DFT exchange-correlation potential
when an electron is added across the gap
\cite{Sham83+}.

The third failure is striking in that it is not easy to see whether it
comes from
the LDA approximation, or from some fundamental feature of DFT.
At first sight, the
dielectric susceptibility, being the second derivative of the
ground state energy with respect to an applied electric
field,
should be obtained exactly within DFT.
Although the corresponding potential change
is linear
in space, and breaks the periodicity of the solid,
techniques were found to bypass this problem
\cite{Baroni86,Levine89}.
Unfortunately, the LDA results
\cite{Baroni86,Levine89,Levine91+}
are usually much larger (more than 10\%) than the experimental data.
Dal Corso, Baroni, and Resta \cite{Dalcorso94} argued that any attempt
to improve the LDA calculated value should go
in the direction of better approximations to the true functional. The
latter
authors tried to compute the dielectric response of Si in the
Generalized
Gradient
Approximation (GGA) \cite{Dalcorso94},
but with limited success.

Earlier, Levine and Allan had
proposed a simple ``LDA + scissor correction'' approach to the
dielectric
tensor\cite{Levine89}, in which a constant shift $\Delta$ is imposed
on the conduction bands with respect to the valence bands, to
reproduce the correct band gap. The dielectric
tensor in this approximation has been found
to be within a few percent of
the experimental data for more than a dozen semiconductors and
insulators
\cite{Levine89,Levine91+}. But,
as emphasized by Dal Corso, Baroni, and Resta, there was no
immediate justification within DFT of this successful procedure.

In recent work by Godby, Sham and Schl\"uter\cite{Godby90},
it was shown that the
exchange-correlation potential near a semiconductor interface
acquires a slow linear spatial variation,
related to the discontinuity, $\Delta$, that occurs on addition of an
electron
to the bulk semiconductor. This variation corresponds to an
ultra-non-local ``vertex correction'' in the Kohn-Sham formulation
of the dielectric response of the semiconductor, and depends  {\it
crucially}
on the electron density at the interface\cite{Godby94}.
A connection between the scissor-operator correction and
this
linear behaviour of the exchange-correlation potential was also pointed
out.

In this paper we reexamine the response of a periodic
solid to an electric field within DFT, and observe that the
original proof of Hohenberg and Kohn \cite{Hohenberg64}
does not apply to this case.
We then prove
the equivalent of the Hohenberg and Kohn theorem in perturbation
theory,
and show that knowledge of both the change in periodic density {\em
and} the change in polarization are needed to recover the change
of periodic potential and change of electric field that induced them.
The exchange-correlation energy thus depends on the
polarization, and this dependence, not taken
into account previously, will generate an exchange-correlation
electric field, identified with the slow linear variation of the
exchange-correlation potential observed by Godby, Sham and
Schl\"uter\cite{Godby90}.
We then explore a
simple model that exhibits the crucial features of our theory,
and explain the relationship to
the scissor-correction approach
to the dielectric tensor.

% DFT
In the DFT approach to the many-body problem, it is shown that
the knowledge of the density $n( {\bf r} )$ of the ground-state  of a
system
with Hamiltonian
$% \begin{equation}
H_v = T + V_{e-e} + v
$ % \end{equation}
(sum of the kinetic $T$, electron-electron
interaction $V_{e-e}$, and one-body local potential $v$
operators), uniquely determines the local potential $v({\bf r})$
of this Hamiltonian, up to a constant.

% Periodic density
Unfortunately,
the straightforward application of an homogeneous electric field
(linear potential), to a system with a periodic potential
does not allow for
a ground-state solution\cite{Nunes94}: a translation against the direction
of the field by a whole number of lattice constants
would always lower the electronic energy. The
impossibility of a ground-state
in the presence of a finite electric field
renders {\em invalid} the original
proof\cite{Hohenberg64} of Density-Functional Theory
for this case.

% Techniques: long-waves or increase of size of a system
The use of perturbation theory allows us to bypass this problem.
We choose to work in the long wave method\cite{Martin72}
(an infinitesimal sinusoidal perturbation whose wavevector
tends to zero). Equivalent results
may be obtained
using other techniques, mentioned in Ref. \cite{Martin72}.
Atomic units, for which
the electronic charge is -1, are used.
The change of total potential corresponding to an infinitesimal
electric
field
$\delta {\cal{E}}$ is (written in one dimension for brevity)
\begin{equation}
\delta v_{\cal{E}}(r)
% = r \, . \, {\delta {\cal{E}}}
  = \lim_{q \rightarrow 0} \delta {\cal{E}} . \frac {\sin (qr)} {q}
  = \lim_{q \rightarrow 0} \frac {\delta {\cal{E}}} {2iq} ( e^{iqr} -
e^{-iqr} ).
\end{equation}

We also allow for changes of potential that are periodic in space,
with the same periodicity as the unperturbed system:
$ %\begin{equation}
\delta v_{G}(r) = \delta v(G) \, . \, e^{iGr}
$ %\end{equation}
with
$ %\begin{equation}
\delta v(G) = (\delta v(-G))^*
$ %\end{equation}
where $G$ is a non-zero vector of the reciprocal lattice.
These changes of potential
are obtained, in the long-wave method, from the Fourier components of
the potential:
\begin{equation}
\delta v_G(r) = \lim_{q \rightarrow 0}
                 \{   \delta v(G \! + \! q) e^{i(G+q)r} +
                      \delta v(G \! - \! q) e^{i(G-q)r}     \}
\end{equation}
such that
$ %\begin{equation}
\frac {\delta v(G)} {2} = \delta v(G \! + \! q)  = \delta v(G \! - \!
q)
$ . %\end{equation}

In response to these perturbations, at finite $q$, the system will
develop
changes in density described similarly by $\delta n(G \! \pm \! q)$.
The
long-wave part
of this change in density, for $q \rightarrow 0$, will
be\cite{Martin72}
\begin{equation}
\delta n_{\cal{P}}(r) =
    \lim_{q \rightarrow 0} \delta {\cal{P}} \, . \, q \sin (qr),
\end{equation}
where $\delta {\cal{P}}$ is the change of polarization for
$q=0$.

The elaboration of a density functional theory for these perturbations
must
answer the following question: what quantities do we need
in order to determine $\delta {\cal{E}}$ and the set of
$\delta
v(G)$ uniquely?

Since we must stay within perturbation theory for treating electric
fields,
we now provide a
perturbative analog of the first Hohenberg-Kohn
theorem\cite{Hohenberg64}:
the knowledge of the {\em change} in density everywhere uniquely
determines
the {\em change} in potential. This theorem could be
proved by taking the infinitesimal limit of finite differences
of the first Hohenberg-Kohn
theorem\cite{Hohenberg64}, but this approach could not be followed
for electric fields, since {\em only} infinitesimal electric fields are
allowed.
The following demonstration stays strictly within perturbation theory.

We consider
the Hylleraas minimum principle \cite{Hylleraas30}:
a trial change in wavefunction $\delta \phi_t$
gives an upper bound on the second-order change in energy
\begin{equation}
\delta^2 E \leq { \langle \delta \phi_t | H-E | \delta \phi_t \rangle
             + ( \langle \delta \phi_t | \delta v | \phi \rangle +
(c.c.) ) }.
\end{equation}
This principle is valid under the constraint
$\langle \delta \phi_t | \phi \rangle + \langle \phi | \delta \phi_t
\rangle = 0$.
The minimum is reached {\em only} for the $\delta \phi$ that is the
response
of the quantum-mechanical system to the change of potential $\delta v$.
When this change of potential is a 1-body local operator, Eq.(4)
becomes
\begin{equation}
\delta^2 E \leq { \langle \delta \phi_t | H-E | \delta \phi_t \rangle
             + \int \delta v(r) \delta n_t(r) dr },
\end{equation}
where the change in density $\delta n_t(r)$
is easily derived from the knowledge of the unperturbed
wavefunction and the trial change in wavefunction.
Now consider two changes in potentials
$\delta v_1(r)$ and $\delta v_2(r)$ such
that $\delta v_1(r) \neq \delta v_2(r)$ + constant.
The Hylleraas minimum principle applied
to the perturbation $\delta v_1(r)$ gives
%\end{multicols}
\begin{equation}
{ \langle \delta \phi_1 | H-E | \delta \phi_1 \rangle
                                                + \int \delta v_1(r)
\delta
n_1(r) dr }
<
{ \langle \delta \phi_2 | H-E | \delta \phi_2 \rangle
      + \int \delta v_1(r) \delta n_2(r) dr },
\end{equation}
%\begin{multicols}{2}[]
while for the perturbation $\delta v_2(r)$,
a similar inequality, where 1 and 2 are interchanged, is obtained.
Summing these two inequalities leads to
\begin{equation}
0
<
\int ( \delta v_1(r) - \delta v_2(r) )  ( \delta n_2(r) - \delta n_1(r)
) dr.
\end{equation}
Setting
$\delta n_1(r) = \delta n_2(r)$ would lead to a contradiction, showing
that two different changes in potential must induce two different
changes in
density.
Thus, the knowledge of $\delta n(r)$ everywhere
uniquely defines the
$\delta v(r)$ that induced it.

The same line of argument can be used in the case of
perturbations of periodic systems with finite wavevector $q$,
as previously defined.
All quantities have to be normalized to the unit cell volume. This
normalization,
and a Fourier
transform, applied to the term
$\int \delta v(r) \delta n(r) dr$ in Eq.(5), changes it into
$\Omega_{cell}
\sum_{G} \{ \delta v^*(G \! + \! q) \delta n(G \! + \! q) +
           \delta v^*(G \! - \! q) \delta n(G \! - \! q) \} $.
The limit $q \! \rightarrow \! 0$ is now taken,
for two different perturbations described
by $\{ \delta {\cal{E}}_1, \delta {v_1}(G) \}$ and
$\{ \delta {\cal{E}}_2, \delta {v_2}(G) \}$. The $G=0$ term is
isolated,
and the
long-wave values from Eq.(1) and (3) are used, such that~:
%\end{multicols}
\begin{equation}
0 <
\frac {\Omega_{cell}} {2}
\,
\{
\, ( \delta {\cal{E}}_1 - \delta {\cal{E}}_2 )
      ( \delta {\cal{P}}_2 - \delta {\cal{P}}_1 )
    +  \sum_{G \neq 0} ( \delta v_1^*(G) - \delta v_2^*(G) )
         ( \delta n_2(G) - \delta n_1(G) )
\,
\}.
\end{equation}
%\begin{multicols}{2}[]
If we now suppose $\delta {\cal{P}}_1=\delta {\cal{P}}_2$ and
$\delta {n_1}(G)=\delta {n_2}(G)$,
the expected contradiction is obtained. From this result we conclude
that the change in potential {\em and} electric field can be deduced from
the knowledge of the change in density {\em and}
polarization that were induced
by them. Note that the knowledge of the change
of polarization is crucial, since it is the quantity
{\em conjugate} to the change of electric field in Eq.(8)~:
if $\delta {\cal{P}}_1$ were allowed to be different from
$\delta {\cal{P}}_2$, Eq.(8) could be satisfied
for some $\delta {\cal{E}}_1 \neq \delta {\cal{E}}_2$.
The dependence on polarization is a remnant of the ultra-non-local
dependence on the long-wave part of the change in density, Eq.(3), that
{\em vanishes} for $q=0$.

Eq.(8) also shows that
{\em if} no change in electric field is allowed, the knowledge
of the polarization is not needed: a
density-polarization functional theory is required for the
{\em specific} case of the response to an electric
field\cite{displacement}.

In the Kohn-Sham construction\cite{Kohn65}, an exchange-correlation
energy
functional
of the density is introduced. Following the Kohn-Sham argument,
now applied to the response to an electric field,
this functional will depend
on both the periodic density {\em and} the polarization:
$ %\begin{equation}
{E_{xc}}[n_{G};{\cal{P}}]
$. %\end{equation}
This functional is
defined only for values of ${\cal{P}}$ in an infinitesimally
small neighborhood of the zero-field polarization\cite{notequantum}.
%Alternatively, it
%can also be expressed in term of the deviation with respect to this
%value:
%$ %\begin{equation}
%{E_{xc}}[n_{G};\delta {\cal{P}}]
%$. %\end{equation}

We show now that this polarization dependence of the
exchange-correlation energy
leads to the existence of a new ``exchange-correlation electric
field''.

First, we mention that the treatment of the electric field as a
perturbation
makes use of the following link between the operator $r$
and the derivative with respect to the wavevector~\cite{Baroni86}:
\begin{equation}
\delta {\cal{E}} \, . \, P_c \, r | u_{j,k} \rangle =
\delta {\cal{E}} \, . \, P_c \, i \, \frac {\partial}
  {\partial k} | u_{j,k} \rangle
\end{equation}
where $P_c$ is the projector on the conduction bands, and
$u_{j,k}$ is the periodic part of the Bloch function
of band $j$ with wavevector $k$.

Also, the recent theory of polarization proposed
by King-Smith and Vanderbilt\cite{King-Smith93+} leads to the
expression
for the polarization\cite{notequantum}
\begin{equation}
{\cal{P}}  =
\sum_{j}  2i \int
%    \left[
     \langle u_{j,k} |  \frac {\partial}
                    {\partial k} |  u_{j,k} \rangle
%+
%           \langle u_{j,k} |  \frac {\partial}
%                    {\partial k} |  \delta u_{j,k} \rangle
%         \right]
      \frac {dk} {2 \pi},
\end{equation}
where the sum is over the valence bands.

When wavefunctions are varied in order to find the Kohn-Sham
self-consistent equations,
the exchange-correlation energy
changes, owing to the implicit dependence of the density on
wavefunctions,
but also owing to the implicit dependence of the polarization on
wavefunctions (see Eq.10).
The term
$ %\begin{equation}
P_c \, \delta v_{xc} | u_{j,k} \rangle
$ %\end{equation}
that appears in the perturbative equivalent of the Kohn-Sham equation
(the Sternheimer equation\cite{Baroni86}) is replaced by
\begin{equation}
P_c \, \delta v_{xc} | u_{j,k} \rangle +
  {\delta \cal E}_{xc} . P_c \, i \, \frac {\partial}
  {\partial k} | u_{j,k} \rangle
\end{equation}
where
$ %\begin{equation}
{\delta \cal E}_{xc} = \frac {\partial^2 E_{xc}} {{\partial\cal P}^2}
\delta {\cal P}
      + \sum_{G} \frac {\partial^2 E_{xc}} {\partial {\cal P} \partial
n_G}
\delta n_G.
$ %\end{equation}

The second term of Eq.(11), missing in previous
theories of the dielectric constant\cite{Baroni86},
is to be interpreted as an additional
electric field (compare with Eq.(9)), created by the polarization
dependence of
the exchange-correlation energy.
It vanishes in zero applied electric field.

We will now analyze the model one-dimensional
semiconductor used by Godby and Sham in Ref.~\cite{Godby94},
and show that it exhibits the main features of our theory:
1) the knowledge of the density change alone is not sufficient to
deduce
the periodic
potential change and electric field change~;
2) an exchange-correlation electric field appears.

In this model, the periodic solid plus electric field
is treated by the long-wave method in a supercell consisting of $N$
basic
unit cells of length {\it a}.
The external potential plus the Hartree potential
$V_{\rm{ext}}(x)+V_{\rm{H}}(x)$
is taken to
be the sum of two different contributions:
$   V_{0} \,  \cos ( \frac{2 \pi x}{a} )
          + \lambda \, \sin ( \frac{2 \pi x}{N a} )
$.
The first term has the periodicity of one unit
cell,
while the second, with the periodicity of the supercell, corresponds to
a slowly varying potential of amplitude $\lambda$. For an infinitely
long
supercell, the second term mimics the action of an electric
field.

The self-energy operator is taken to be the non-local potential:
$ %\begin {equation}
 \Sigma (x,x',\omega)=
  \frac {f(x)+f(x')}{2} g(|x-x'|)
$ %\end {equation}
where $f(x)=-F_{0} [1-cos(2 \pi x /a)]$ is a negative function
with the periodicity of one unit cell and $g(y)$ is a normalized
gaussian of width $w$ = 2 a.u.

% Results.
We keep the same set of parameters as in Ref.~\cite{Godby94}.
First, the many-body problem is solved by direct diagonalization
of the equations containing the self-energy operator, using
a plane-wave basis set and a Brillouin zone sampling at the $\Gamma$
point only. From this result, we construct an {\em exact} density
functional
theory by determining the local potential $V_{\rm eff,1}(x)$ which,
when
filled
with {\em non-interacting} electrons (no self-energy operator),
reproduces the same electron density
as in the many-body case (see Fig. 4 of Ref.~\cite{Godby94}).
Standard iterative non-linear optimization techniques
are used for that purpose.
Independently,
we also extract the (polarized) density of the single unit cell
$\pm  \frac{1}{2} a$
where the slowly varying potential is most linear
and, again using optimization techniques,
we reproduce this density with {\em another} potential,
$V_{\rm eff,2}(x)$, having period {\it a}.
In this case, to retain the sampling of the Brillouin zone
at the $\Gamma$ point, we repeat the density periodically in the
supercell.
The only problem in reproducing the ``target'' density comes from
the slight non-linearity of the slowly varying potential, and
disappears progressively for larger supercells.
Fig. 1 presents the results
obtained with a 80{\it a} supercell and $\lambda$= 0.95 eV,
for which the r.m.s. difference between the target and
computed density is 4 parts per thousand.

% conclusions of the model.
This example illustrates that the action of a linear potential
$\Delta (V_{\rm ext}+V_{\rm H})$, superimposed on a periodic potential, in the
case of an interacting-electron model system
can be mimicked by {\em two} different potentials
$\Delta
V_{\rm eff,1}$ and $\Delta
V_{\rm eff,2}$, acting upon
the corresponding non-interacting electron system,
in contrast to a na{\"{\i}}ve application of DFT theorems.
However, $\Delta
V_{\rm eff,2}$ is not able to
reproduce the polarization (0.037 electrons),
associated with the long-wavelength charge density.
Moreover, $\Delta (V_{\rm ext}+V_{\rm H})$
and the linear component $\Delta V^{\rm linear}_{{\rm
eff,1}}$ of the Kohn-Sham potential $\Delta V_{{\rm
eff,1}}$
differ by 15\%, owing to the existence of the exchange-correlation
electric field.

We now discuss briefly the implications of our theory for practical
calculations.
LDA or GGA techniques can be used to
approximate the periodic-density dependence of $E_{xc}$.
Unfortunately, since the homogeneous electron gas is metallic,
we do not see any easy way to modify LDA or
GGA in order to obtain a polarization-dependent
exchange-correlation energy, that would generate the second term in
Eq.(11).
Nevertheless, our theory allows the
scissor-operator correction to be understood as
an {\em approximate} way of {\em bypassing} this dependence.

Indeed, in Ref.~\cite{Godby94}, it was shown that
the effect of a slowly varying potential
$\Delta V$ that induces transitions
between valence and conduction bands with a gap $E_g$ (see Eq.(10) of
Ref.~\cite{Godby94}) is approximately equivalent to the effect of
a slowly varying potential
$\Delta V + \Delta V_{xc}$ that induces transitions
between valence and conduction bands with a gap $E_g^{DFT}$ (see
Eq.(11) of
Ref.~\cite{Godby94}). Hence, the {\em exact} density-polarization
functional theory calculation in a field
$\delta {\cal{E}} + \delta {\cal{E}}_{xc}$, with a gap $E_g^{DFT}$, can
be
{\em approximately} replaced
by a LDA calculation in a field $\delta {\cal{E}}$, with a
modified gap $E_g$: this is
the ``LDA + scissor correction'', which gives a dielectric
constant for Si of 11.2 (Experiment 11.4-11.7, LDA 12.9-13.5, GGA
12.4-12.6) \cite{Levine89,Levine91+,Dalcorso94}.

In conclusion, because the original Hohenberg-Kohn theorem
is not valid for the case of the response to an homogeneous electric field,
we have set up
a density-polarization functional theory, of
which an approximate form is the ``LDA + scissor correction''.

%\acknowledgments

The authors acknowledge interesting discussions or comments from
R. Resta, D. Vanderbilt, A. Dal Corso, F. Mauri, and M.S.
Hybertsen, and
are grateful to FNRS-Belgium (X.G. and Ph.G),
IRSIA-Belgium, the British Council
(Ph.G.),
The Royal Society,
EPSRC and the EU programme
Human Capital and Mobility [contract no. CHRX-CT93-0337] (R.W.G.),
for financial support.
%\vspace{2mm}

%\begin{thebibliography}{10}

\begin{center}
\section*{Figure Captions}
\end{center}

\begin{enumerate}

\item[Fig.1]
The unit cell of our model one-dimensional
semiconductor, where the slowly varying applied potential
$\Delta (V_{\rm ext}+V_{\rm H})$, that changes the
interacting electron density by $\Delta n$,
is the most linear.
Both $\Delta V_{{\rm eff,1}}$ and  $\Delta V_{\rm eff,2}$,
used in the non-interacting Kohn-Sham equations,
yield the same $\Delta n$.
$\Delta V_{\rm eff,2}$ is a periodic potential with no
linear slope,
while $\Delta V_{{\rm eff,1}}$, whose linear part is $\Delta V^{\rm
linear}_{{\rm
eff,1}}$,
reproduces not only $\Delta n$ but also the change of polarization due
to
$\Delta (V_{\rm ext}+V_{\rm H})$.
This illustrates the need for polarization-dependence in $E_{xc}$.
For clarity, the potential curves have been aligned
so that they all start from zero.

\end{enumerate}

%\end{multicols}


\begin{references}
\bibitem{displacement}
As usual in the theory of electromagnetism,
the electric displacement field
has to be used instead of the electric field
if the electron-electron interaction is not short-ranged, which is the
case
for the Coulombic interaction. See
L.D. Landau and E.M. Lifshits, Electrodynamics of Continuous
Media (Pergamon Press, 1960). All the results of our theory
are unaffected by this replacement.
\bibitem{notequantum}
The modern theory of polarization\cite{King-Smith93+}
defines it only up to a quantum.
This feature is of no consequence for the present theory,
since only infinitesimal changes in polarization are considered.
\end{references}
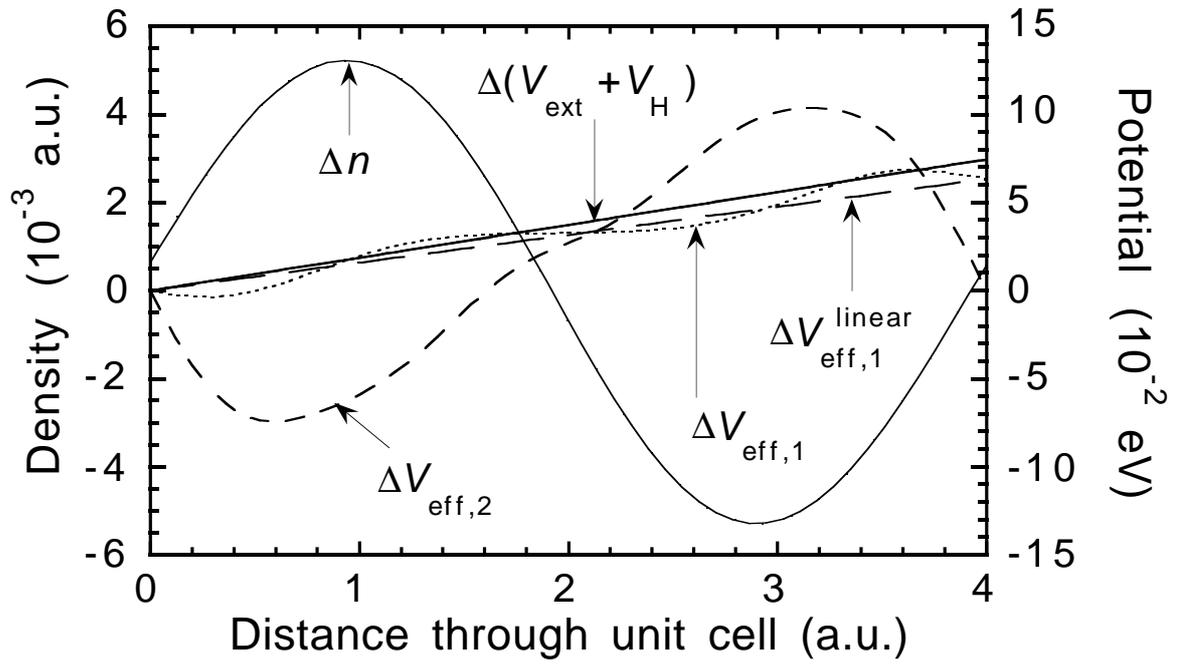
\end{document}